\documentclass{appolb}
\usepackage{graphicx}
\usepackage{amsmath}

\begin{document}
\title{Timescales in heavy ion collisions
\thanks{Written for the symposium in honor of Andrzej Bia\l as  $80^{\rm th}$ birthday}%
}
\author{Mike Lisa
\address{Department of Physics, The Ohio State University}
}
\maketitle
\begin{abstract}
The study of high energy collisions between heavy nuclei is a field unto itself, distinct from nuclear and particle physics.
A defining aspect of heavy ion physics is the importance of a bulk, self-interacting system with a rich space-time substructure.
I focus on the issue of timescales in heavy ion collisions, starting with proof from low-energy collisions that
  femtoscopy can, indeed, measure very long timescales.
I then discuss the relativistic case, where detailed measurements over three orders of magnitude in energy reveal
  a timescale increase that might be due to a first-order phase transition.
I discuss also consistency in evolution timescales as determined from traditional longitudinal sizes and a novel
  analysis using shape information.
\end{abstract}

\begin{flushleft}
\begin{small}
    The slowly crawling ants will eat our dreams.\\\textit{- Andrzej Bia\l as, musing on words of Andre Breton}\\
\medskip
    Go to the ant, thou sluggard; consider her ways, and be wise.\\\textit{- Proverbs vi.6}
\end{small}
\end{flushleft}

\section{Preface}

In the quote above, made at the first Workshop on Particle Correlations and Femtoscopy
  in the Czech Republic, Professor Bia\l as was expressing a frustration felt periodically by
  those of us who labor to understand deeply the fascinating features of soft-scale QCD
  as manifest in the quark-gluon plasma, a bulk thermodynamic system of deconfined colored
  partons as degrees of freedom.
Every time we gain a deeper insight into the physics and phenomenology of this system (the dream),
  more detailed theories (the ants) or experimental observations make clear that the system is more complicated
  than we thought.
New advances often raise more questions than they answer.

Professor Bia\l as made this statement with a smile on his face, however.
He clearly considers himself an ant in the spirit of the quote from Proverbs: a worker
  with a mission much larger than himself, destined to build, piece by piece over the course
  of his life, an edifice in pursuit of that mission.
He clearly relishes this role.

It turns out that this symposium is held shortly before a milestone birthday of my own,
  and I found myself contemplating my own much less impressive anthills.
One topic I have returned to repeatedly in various forms is the timescale of the system
  formed in a heavy ion collision.
Here, I discuss previous studies (and one unpublished analysis) to show the development
  of our understanding of these timescales as measured with two-particle intensity interferometry.



\section{Introduction}

To the general public, the field of heavy ion physics resembles high energy particle physics.
The accelerators, the collaborations, and the detectors are mammoth.
Papers are written by committee, and talks are selected according to the bylaws set
  by Councils and led by elected management teams.
Students are well-versed in the particle zoo (often much more so, than their professors
  who grew up as nuclear physicists).

The origins of the field, however, lie more in the realm of nuclear physics.
Concepts were developed and people trained in heavy ion experiments at facilities
  like GANIL, SIS/GSI, and the NSCL/MSU cyclotron facility.
Pions were rare and almost exotic.
Students were relatively well-versed in nuclear physics.

However, heavy ion physics is a field of its own-- neither nuclear physics (which strives
  to understand the nucleus in its seemingly infinite complexity) nor particle physics
  (which attempts to bypass the complexity of all interactions to study symmetries 
  manifest as particles).
In heavy ion physics, we seek to create and study a new {\it system}.
Ideally, it will be a nearly thermalized system, so that we may study its equation of
  state.
At lower energies, the equation of state of highly compressed, cold matter provides
  information relevant to the cores of neutron stars~\cite{Tsang:2011ju}.
In ultrarelativistic energies, the equation of state of colored matter near the
  deconfinement transition probes QCD under the most extreme conditions~\cite{Shuryak:2008eq}.

\begin{figure}[t]
{\centerline{\includegraphics[width=0.8\textwidth]{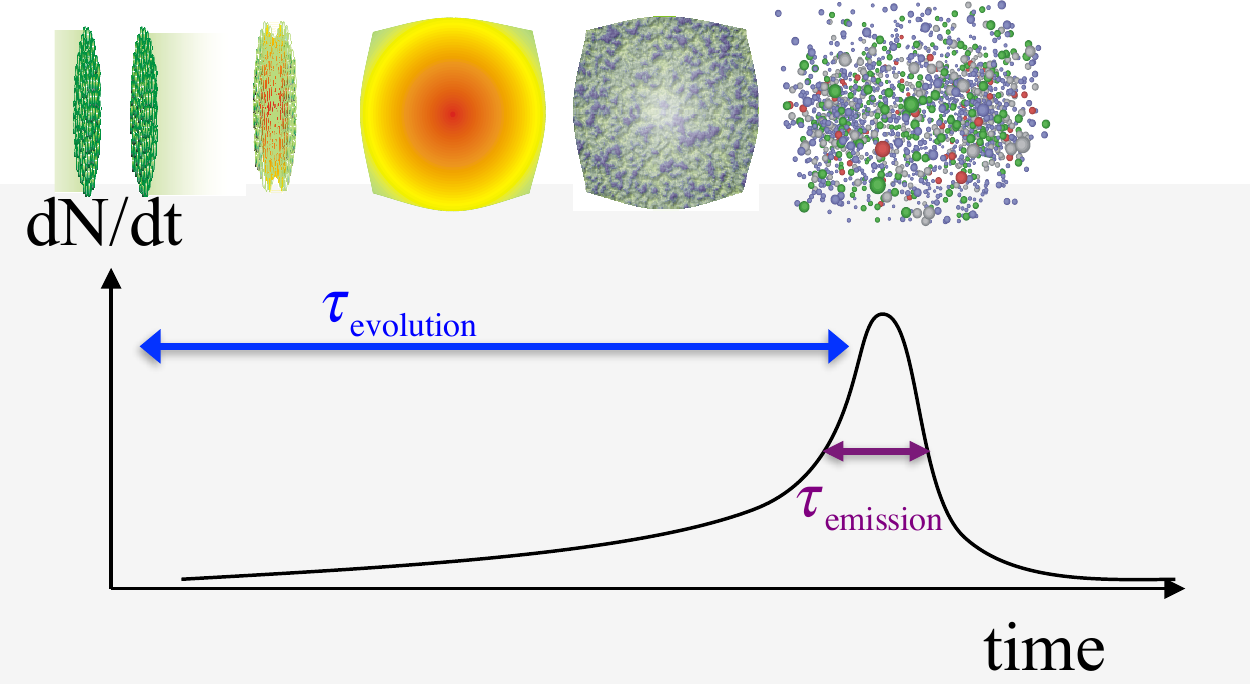}}}
\caption{
\label{fig:EvolutionCartoon}
The evolution of an ultra-relativistic heavy ion collision is sketched to indicate two
  relevant timescales, corresponding to the evolution of the entire system and the
  duration of the freezeout process.
See text for details.}
\end{figure}

The hot system is self-interacting and characterized by detailed flow fields.
Its femtoscopic substructure is dynamic and rich, with long lifetimes, anisotropic
  shapes, correlations between momentum and space-time, and whirling vortices.
To understand the evolution of this substructure, it is important to obtain measures
  of the timescales involved.
Figure~\ref{fig:EvolutionCartoon} identifies two of them, for the case of ultrarelativistic collisions.
Unfortunately, they are often conflated, using the ambiguous term ``lifetime;'' however, they are
  different, and it is best to keep the distinction clear.
The evolution timescale $\tau_{\rm evolution}$ refers to the time between initial interpenetration
  and particle freezeout.
(Particles ``freeze out'' when they cease interacting with each other and the system.)
Meanwhile, $\tau_{\rm emission}$ refers to the duration of the freezeout process itself.

\section{Can femtoscopy measure long emission durations?}

The technique of two-particle intensity interferometry is a well-developed tool to extract spatio-temporal
  information from dynamic subatomic sources.
Also known as femtoscopy, it exploits the fact that, given the observation of one particle, the conditional
  probability to measure a second particle depends on the relative momentum (measured) and the relative space-time
  position (inferred, by measuring the conditional probability) of the pair.
For details and compilations of results, I refer the reader to reviews at both low~\cite{Boal:1990yh}
  and high~\cite{Lisa:2005dd} energy collisions.

In principle, information about both space and time scales may be extracted by 
  studying multi-dimensional correlation functions in the ``out-side-long'' (or, for low energies,
  the ``longitudinal-transverse'') system of Bertsch and Pratt~\cite{Pratt:1986cc,Bertsch:1989vn}.
Here, the ``out'' (or, for low energies, the ``longitudinal'') direction is parallel to the direction
  of motion of the particles, while the ``side'' (or ``transverse'') is perpendicular to it.
A long emission duration ($\tau_{\rm emission}$) will generate a particle distribution
  extended in the direction of particle motion, and the resulting correlation
  will be less if the relative momentum is oriented in this direction.
Emission duration measurements are of particular interest, because a first-order phase transition from a deconfined
  to a confined state, is expected to extend the emission time~\cite{Pratt:1986cc,Bertsch:1989vn,Rischke:1996em}.

Through the early nineties, no emission duration greater than $\sim$2~fm/c had been observed in
  the correlation data; indeed, most extracted timescales were consistent with zero.
For the newly available collisions at the multi-GeV scale~\cite{Lisa:2005dd}, this was a disappointing
  development, though perhaps not shocking.
However, at non-relativistic energies available at NSCL and GANIL, this was surprising indeed.
At these lower energies, the collision and evolution dynamics were believed to be better understood.
Repeated reports of vanishing timescales from correlation measurements led some experts at the time
  to wonder whether the femtoscopic technique itself was sufficiently well
  understood~\footnote{Scott Pratt, 1992, private communication}.

Could femtoscopy really measure timescales, after all?
Two publications~\cite{Lisa:1993zz,Lisa:1994zz} on proton correlations answered this important question with a resounding affirmative.
In one, near multifragmentation energies, a lifetime greater than 10~fm/c was finally extracted from femtoscopic
  data.
In the other, at compound nucleus energies, a lifetime greater than 1000~fm/c (!!) was reported.
Both timescales were of the order of theoretical expectations.

Why had all previous published results reported no difference between longitudinal and transverse correlation functions,
  and hence emission timescales consistent with zero?
The reason turned out to be simple: at least in the U.S., we
  had all been looking in the wrong frame.

\subsection{A study of collisions at ``intermediate'' energies}
\label{sec:ArSc80}

Two-proton correlation functions at small relative momenta probe the space-time geometry of the emitting system, 
  because the magnitude of nuclear and Coulomb
  final-state interaction and antisymmetrization effects depends on the spatial separation of the emitted particles~\cite{Koonin:1997fh}. 
The attractive S-wave nuclear interaction leads
  to a pronounced maximum in the correlation function at
  relative momentum $q=20$~MeV/c. 
This maximum decreases for increasing source dimensions and/or emission
  time scales. 
The Coulomb interaction and antisymmetrization produce a minimum at $q=0$.
Nonspherical phase-space distributions, predicted for long-lived emission sources, 
  can lead to a dependence of the two-proton correlation function on the direction of the relative
  momentum~\cite{Pratt:1987zz}. 
Until 1993, however, such directional dependences had not yet been observed
  unambiguously.
The first observation was published in 1993~\cite{Lisa:1993zz}.

The experiment was performed at the National Superconducting Cyclotron Laboratory at Michigan
  State University (MSU). 
A beam of Ar ions at E/A=80 MeV
  incident energy and intensity $\sim3\times10^8$/sec
  bombarded a Sc target of areal density 10~mg/cm. 
Charged particles were measured in the
MSU $4\pi$ Array, which consisted of 209 plastic $\Delta E-E$
  phoswich detectors covering polar angles between $7^\circ-158^\circ$
  in the laboratory frame.
One of the hexagonal modules of the $4\pi$ Array, located at
  $38^\circ$ in the laboratory frame, was replaced by a 56-element high-resolution hodoscope.
Each $\Delta E-E$ telescope of the hodoscope consisted of a $300-\mu{\rm m}$-thick Si detectors backed by
  a 10-cm-long CsI(Tl) detector and subtended a solid angle of $\Delta\Omega=0.37$~msr.
The energy resolution was about 1\% for 60~MeV protons; this is important for measuring
  large source sizes.

\begin{figure}[t]
{\centerline{\includegraphics[width=0.5\textwidth]{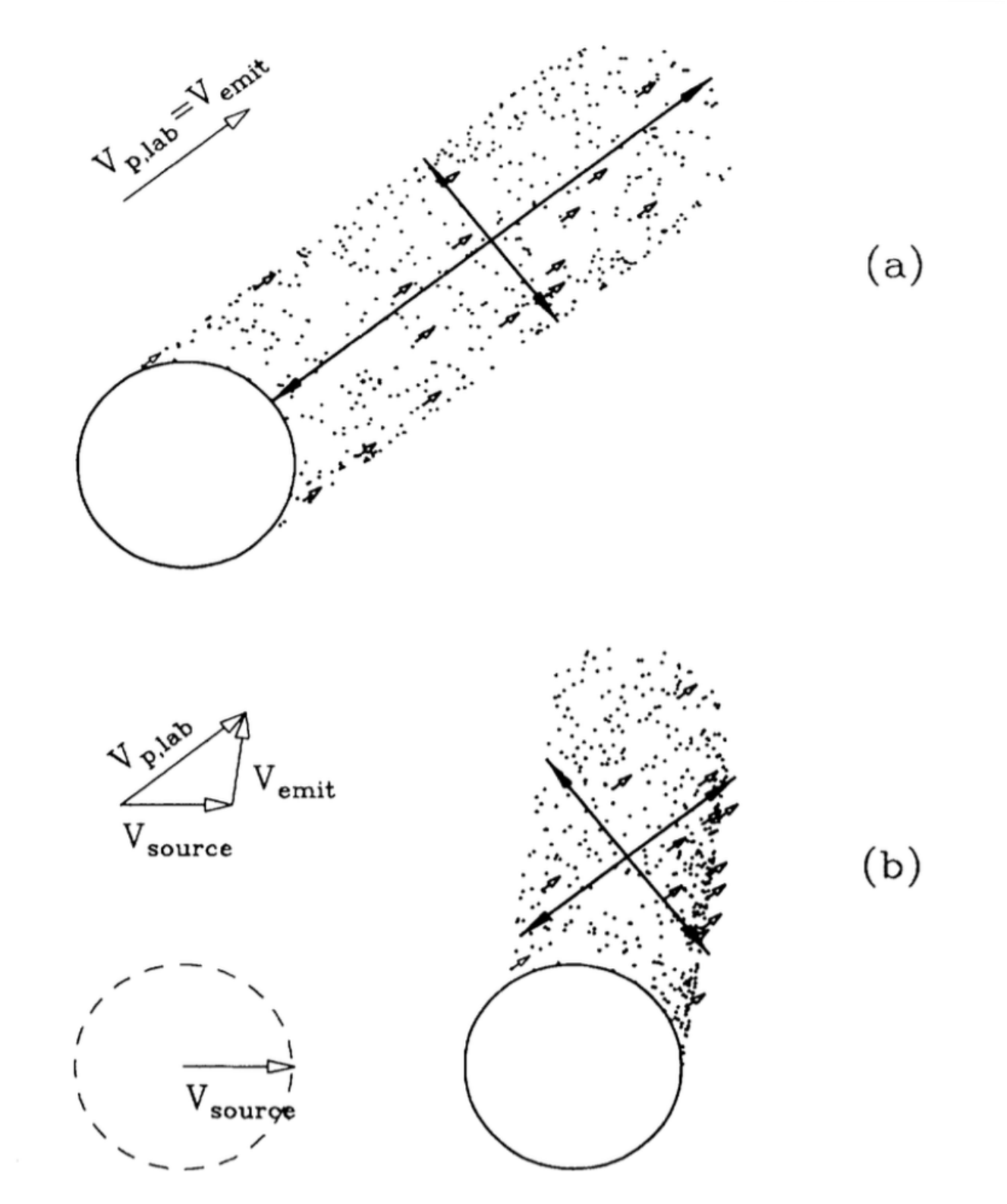}}}
\caption{
\label{fig:Lisa1993fig1}
Schematic illustration of phase-space distributions at a time $t=70$~fm/c, seen by a detector
  at $\theta_{\rm lab}=38^\circ$, for a spherical source of radius $r=3.5$~fm and lifetime $\tau=70$~fm/c
  emitting protons of momentum 250~MeV/c.
(a) Source at rest in the laboratory.
(b) Source moves with $v_{\rm source}=0.18c$.
In the phase-space distributions, the laboratory velocities of the emitted particles
  ($\vec{v}_{\rm p,lab}$) are depicted by small arrows, and the directions perpendicular
  and parallel to $\vec{v}_{\rm p,lab}$ are depicted by the large double-headed arrows.
In (a) and (b), $\vec{v}_{\rm p,lab}$ is kept constant, and $\vec{v}_{\rm emit}$ is
  different; therefore, the elongations along $\vec{v}_{\rm emit}$ are different.
From~\cite{Lisa:1993zz}.
}
\end{figure}

The problem of identifying finite emission duration is illustrated in
  Figure~\ref{fig:Lisa1993fig1}.
It depicts phase-space distributions in the laboratory rest frame of protons
  emitted with fixed laboratory velocity $\vec{v}_{\rm p,lab}$ towards the detector
  at $\theta_{\rm lab}=38^\circ$ for a source at rest in the laboratory (a) and
  for a source at rest in the center-of-momentum system of the projectile and target
  ($v_{\rm source}=0.18c$).
We assumed a spherical source of 7~fm diameter and 70~fm/c lifetime emitting protons
  of momentum 250~MeV/c.

\begin{figure}[ht]
{\centerline{\includegraphics[width=0.5\textwidth]{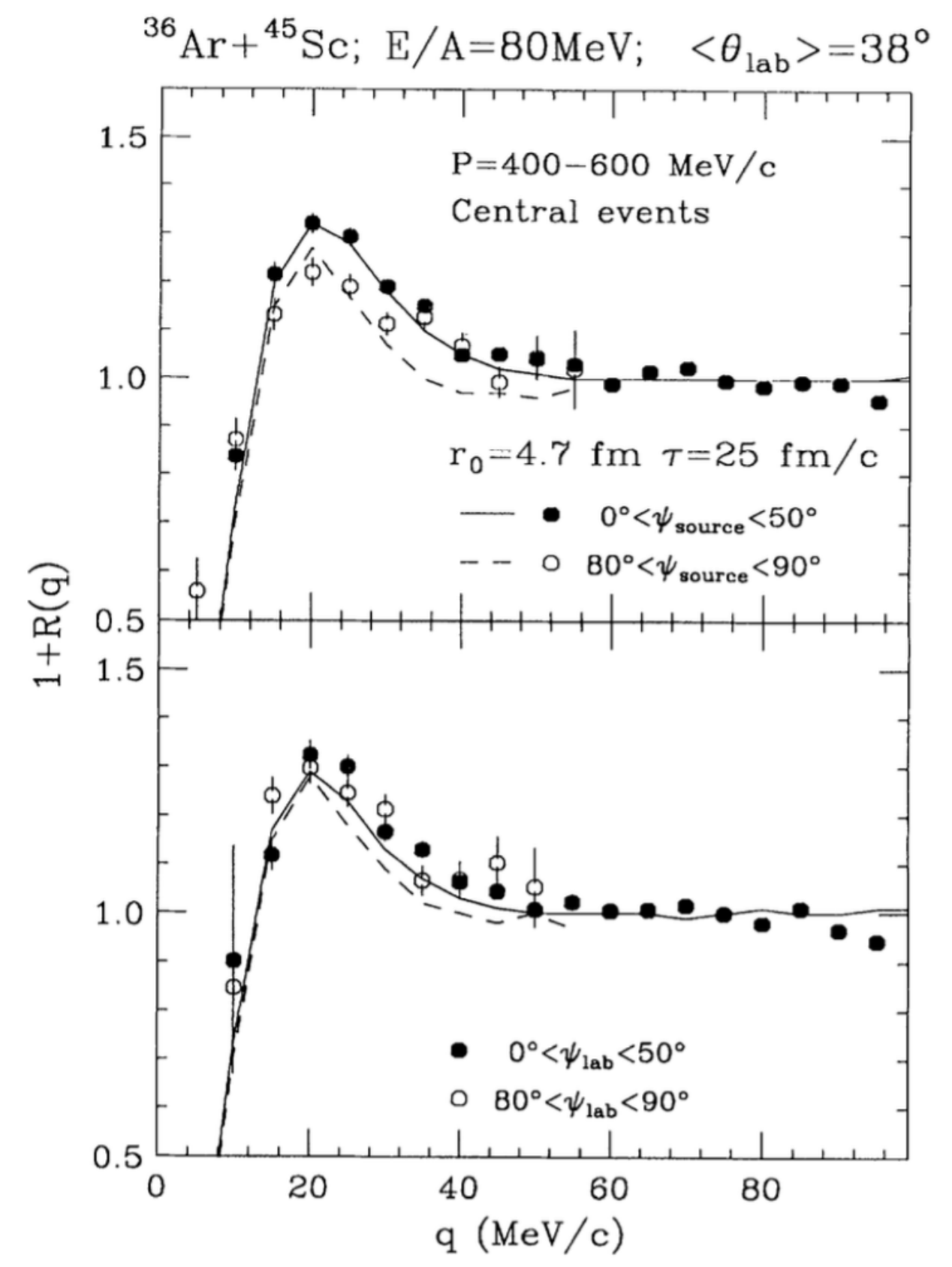}}}
\caption{
\label{fig:Lisa1993fig2}
Measured longitudinal and transverse correlation functions for protons emitted in central $^{36}{\rm Ar}+^{45}{\rm Sc}$
  collisions at $E/A=80$~MeV.
The correlation functions are shown for proton pairs of total laboratory momentum $P=400-600$~MeV/c
  detected at $\langle\theta_{\rm lab}\rangle=38^\circ$.
Longitudinal and transverse correlation functions (solid and open points, respectively) correspond to 
  $\psi=\cos^{-1}\left(\vec{q}\cdot\vec{P}/qP\right)=0^\circ-50^\circ$ and $80^\circ-90^\circ$, respectively.
Solid and dashed curves represent longitudinal and transverse correlation functions predicted for emission from 
  a spherical Gaussian source with $r_0=4.7$~fm and $\tau=25$~fm/c, moving with $v_{\rm source}=0.18c$.
Upper panel: $\vec{P}$ and $\psi$ are defined in the rest frame of the presumed source.
Lower panel: $\vec{P}$ and $\psi$ are defined in the laboratory frame.
From~\cite{Lisa:1993zz}.
}
\end{figure}

For emission from a source at rest, the phase-space distribution of particles moving
  with fixed velocity $\vec{v}_{\rm p,lab}=\vec{v}_{\rm emit}$ towards the detector
  exhibits an elongated shape oriented parallel to $\vec{v}_{\rm p,lab}$.
A source of lifetime $\tau_{\rm emission}$ appears elongated in the direction of the proton momentum
  by an incremental distance 
  $\Delta\vec{s}\approx\vec{v}_{\rm emit}\cdot\tau_{\rm emission}=\vec{p}_{\rm p,lab}$.
Correlation functions for relative momenta $\vec{q}\perp\vec{v}_{\rm p,lab}$ reflect
  a stronger Pauli suppression, and hence a reduced maximum at $q\approx20$~MeV/c, than
  those for $\vec{q}\parallel\vec{v}_{\rm p,lab}$.

Cuts on the relative orientation of $\vec{q}$ and $\vec{P}$ are sensitive to the motion of
  the source, since the direction of the total momentum depends on the rest frame, while
  the direction of the relative momentum-- at least in the nonrelativistic limit-- does
  not.
(Note: the key to this result is not even a relativistic boost, but simply a Galilean one!)
Previous analyses compared the shapes of the correlation functions selected
  by cuts on the relative angle $\psi_{\rm lab}=\cos^{-1}\left(\vec{q}\cdot\vec{P}/qP\right)$
  between $\vec{q}$ and $\vec{P}=\vec{p}_1+\vec{p}_2\approx2m\vec{v}_{\rm p,lab}$,
  where $\vec{p}_1$ and $\vec{p}_2$ are the laboratory momenta of the two protons and
  $\vec{q}$ is the momentum of relative motion.
Such analyses are optimized to detect emission duration effects of sources stationary in the laboratory
  system, but they can fail to detect such effects for nonstationary sources.
For the specific case illustrated in Fig.~\ref{fig:Lisa1993fig1}(b), the source dimensions
  parallel and perpendicular to $\vec{p}_{\rm p,lab}$ are very similar, and no significant
  differences are expected for the corresponding longitudinal and transverse correlation
  functions.

For a source of known velocity, the predicted lifetime effect is detected most
  clearly if longitudinal and trans- verse correlation functions are selected 
  by cuts on the angle $\psi_{\rm source}=\cos^{-1}\left(\vec{q}^\prime\cdot\vec{P}^\prime/q^\prime P^\prime\right)$,
  where the primed quantities 
  are defined in the rest frame of the source. 
In the frame of the source, the phase-space distribution is always elongated 
  in the direction of $\vec{v}_{\rm emit}$.
Hence, in Fig.~\ref{fig:Lisa1993fig1}(b), the source dimensions should 
  be compared in directions parallel and perpendicular to $\vec{v}_{\rm emit}$.
Such analyses can only be carried out for emission from well-characterized sources.

Figure~\ref{fig:Lisa1993fig2} corroborates this reasoning with experimental data.
It shows longitudinal and transverse two-proton correlation functions for central
  Ar+ Sc collisions at E/A=80 MeV selected by appropriate cuts on the total
  transverse energy detected in the $4\pi$ Array. 
In a geometrical picture, the applied cuts correspond to reduced impact 
  parameters of $b/b_{\rm max}=0-0.36$. 
Longitudinal (solid points) and transverse (open points) correlation functions
  were defined by cuts on the angle $\psi=\cos^{-1}\left(\vec{q}\cdot\vec{P}/qP\right)=0^\circ-50^\circ$
  and $80^\circ-90^\circ$, respectively.
The normalization constant C in Eq. (1) is independent of $\psi$.
To maximize lifetime effects and reduce contributions from the very early stages 
  of the reaction, the coincident proton pairs were selected by a low-momentum cut  
  on the total laboratory momentum, $P=400-600$~MeV/c. 
The top panel shows correlation functions for which the angle $\psi$ was defined 
  in the center-of-momentum frame of projectile and target ($\psi=\psi_{\rm source}$);
  for central collisions of two nuclei of comparable mass, this rest frame should be close
  to the rest frame of the emitting source.
The bottom panel shows correlation functions for which the angle $\psi$ was defined
  in the laboratory frame.

Consistent with the qualitative arguments presented in Figure~\ref{fig:Lisa1993fig1},
  a clear difference between longitudinal and transverse correlation functions is
  observed for cuts on $\psi_{\rm source}$ (top panel of figure~\ref{fig:Lisa1993fig2})
  but not for cuts on $\psi_{\rm lab}$ (bottom panel of figure~\ref{fig:Lisa1993fig2}).
The clear suppression of the transverse
  correlation function with respect to the longitudinal
  correlation function observed in the top panel in Figure~\ref{fig:Lisa1993fig2} is
  consistent with expectations for emission from a source of
  finite lifetime.
The solid and dashed curves
  in the top and bottom panels of Figure~\ref{fig:Lisa1993fig2} depict calculations
  for emission from a spherical Gaussian source comoving
  with the center-of-momentum frame of the projectile and
  target. The calculations were performed for the radius
  and lifetime parameters $r_0=4.7$~fm and $\tau_{\rm emission}=25$~fm/c
The calculations corroborate the qualitative arguments illustrated 
  in Figure~\ref{fig:Lisa1993fig1}. 
The data in Figure~\ref{fig:Lisa1993fig2} represent the first clear experimental evidence of
  this predicted lifetime effect.


For a more quantitative analysis, we performed calculations assuming a simple
  family of sources of lifetime ~ and spherically symmetric Gaussian density profiles,
  moving with the center-of-momentum frame of reference. 
Energy and angular distributions of the emitted protons were selected by 
  randomly sampling the experimental yield $Y(\vec{p})$. 
Specifically, the single 
  particle emission functions were parametrized as
\begin{equation}
g\left(\vec{r},\vec{p}t\right)\propto\exp\left(-r^2/r^2_0-t/\tau\right)Y\left(\vec{p}\right) . \label{eq:GaussSource1}
\end{equation}

\begin{figure}[t]
{\centerline{\includegraphics[width=0.7\textwidth]{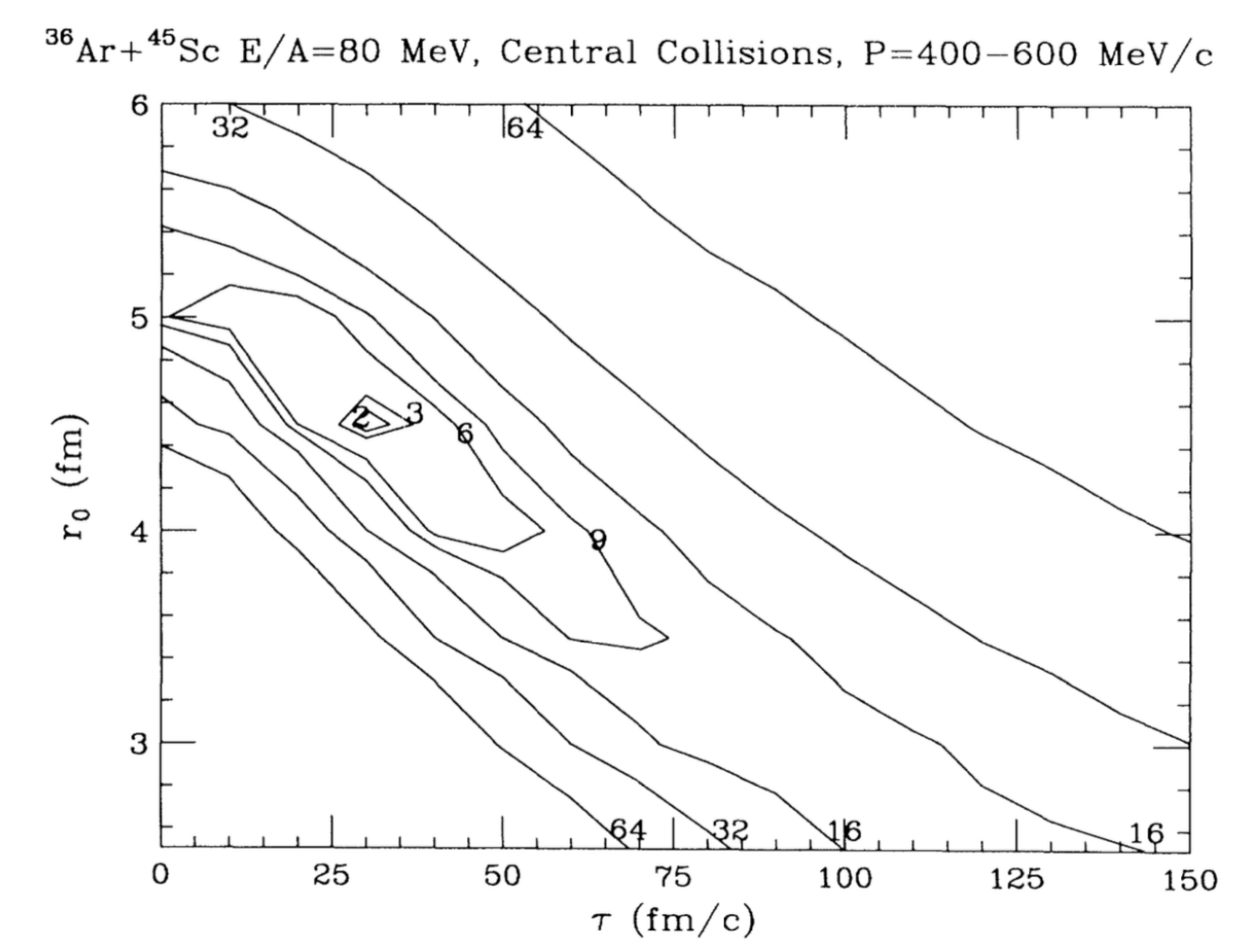}}}
\caption{
\label{fig:Lisa1993fig3}
Contour diagram of $\chi^2/\nu$ determined by comparing theoretical correlation functions
  to the data shown in the upper panel of figure~\ref{fig:Lisa1993fig2}.
The fit was performed in the peak region of the correlation function $q=15-30$~MeV/c.
From~\cite{Lisa:1993zz}.
}
\end{figure}

In equation~\ref{eq:GaussSource1}, $\vec{r}$, $\vec{p}$, and $t$ are understood as being in
   the rest frame frame of the source.
Phase-space points generated in the rest frame of the source were Lorentz boosted into the
  laboratory frame, and the two-proton correlation function was obtained by convolution
  with the two-proton relative wavefunction.

Transverse and longitudinal correlation functions were calculated for the range
  of parameters $r_0=2.5-6.0$~fm and $\tau=0-150$~fm/c. 
For each set of parameters, the agreement between calculated and measured 
  longitudinal and transverse correlation functions was evaluated by
  determining the value of $\chi^2/\nu$ in the peak region, $q=15-30$~MeV/c.
A contour plot of $\chi^2/\nu$ as a function of $r_0$ and $\tau$ is given in
  figure~\ref{fig:Lisa1993fig3}.
Good agreement between calculations and data is obtained for source parameter values
  of roughly $r_0\approx4.5-4.8$~fm and $\tau\approx20-40$~fm/c. 
These extracted emission time scales are qualitatively consistent with those predicted
  by microscopic transport calculations.

\subsection{Very long emission durations from Xe+Al collisions}
\label{sec:XeAl31}

The measurement discussed in the previous section provided the first unambiguous
  observation of long emission durations with femtoscopy.
It thus validated the technique-- source lifetimes (emission durations) {\it can} be measured.
For years, the problem had been that we were looking at longitudinal and transverse
  cuts in the wrong (laboratory) frame.

Dynamical models for symmetric systems with beam energies $E/A\approx80$~MeV predict
  lifetimes $\sim20$~fm/c, consistent with data, as we've seen.
But {\it really} long lifetimes are predicted at lower excitation energies, where
  a compound nucleus is briefly formed and cools by nucleon emission.

A study of two-proton correlation functions in the inverse kinematics reaction Xe+Al at $E/A=31$~MeV, 
  reported~\cite{Gong:1991zz} no difference between longitudinal and transverse correlation functions, 
  although a very long lifetime ($\tau~\sim1000$~fm/c) would be expected.
With the newfound insight on the importance of analyzing the data in the ``right'' (source) frame,
  we decided to extract the raw data from storage and perform a re-analysis.

\begin{figure}[t]
{\centerline{\includegraphics[width=0.7\textwidth]{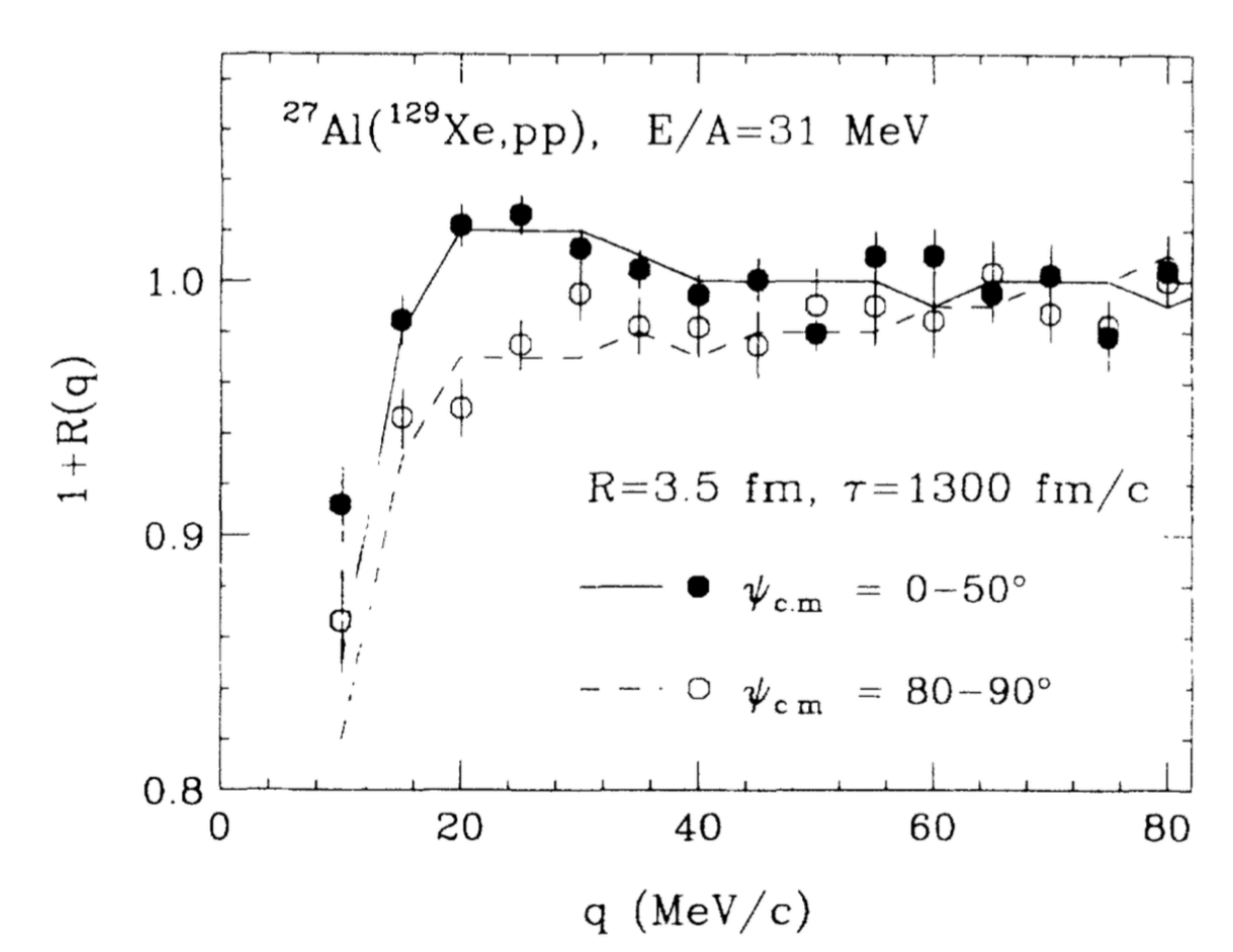}}}
\caption{
\label{fig:Lisa1994fig3}
Comparison of measured (points) and calculated (curves) correlation functions.
The calculations were performed for emission from a schematic source with radius and
  lifetime parameters $R=3.5$~fm and $\tau=1300$~fm/c.
From~\cite{Lisa:1994zz}.
}
\end{figure}

The results are shown in figure~\ref{fig:Lisa1994fig3}.
When we repeated the analysis of~\cite{Gong:1991zz}, we found no difference when cutting on $\psi_{\rm lab}$,
  in agreement with the original published result.
However, when we selected on the angle between $\vec{q}$ and $\vec{P}$ in the center-of-mass frame, 
  a significant difference was observed~\cite{Lisa:1994zz}.
The curves in figure~\ref{fig:Lisa1994fig3} correspond to a spherical source, moving in the
  lab at $v_{\rm source}=0.2086c$ (the system center-of-momentum velocity), with radius and
  lifetime parameters $R=3.5$~fm/c and $\tau=1300$~fm/c, respectively.

\begin{figure}[t]
{\centerline{\includegraphics[width=0.7\textwidth]{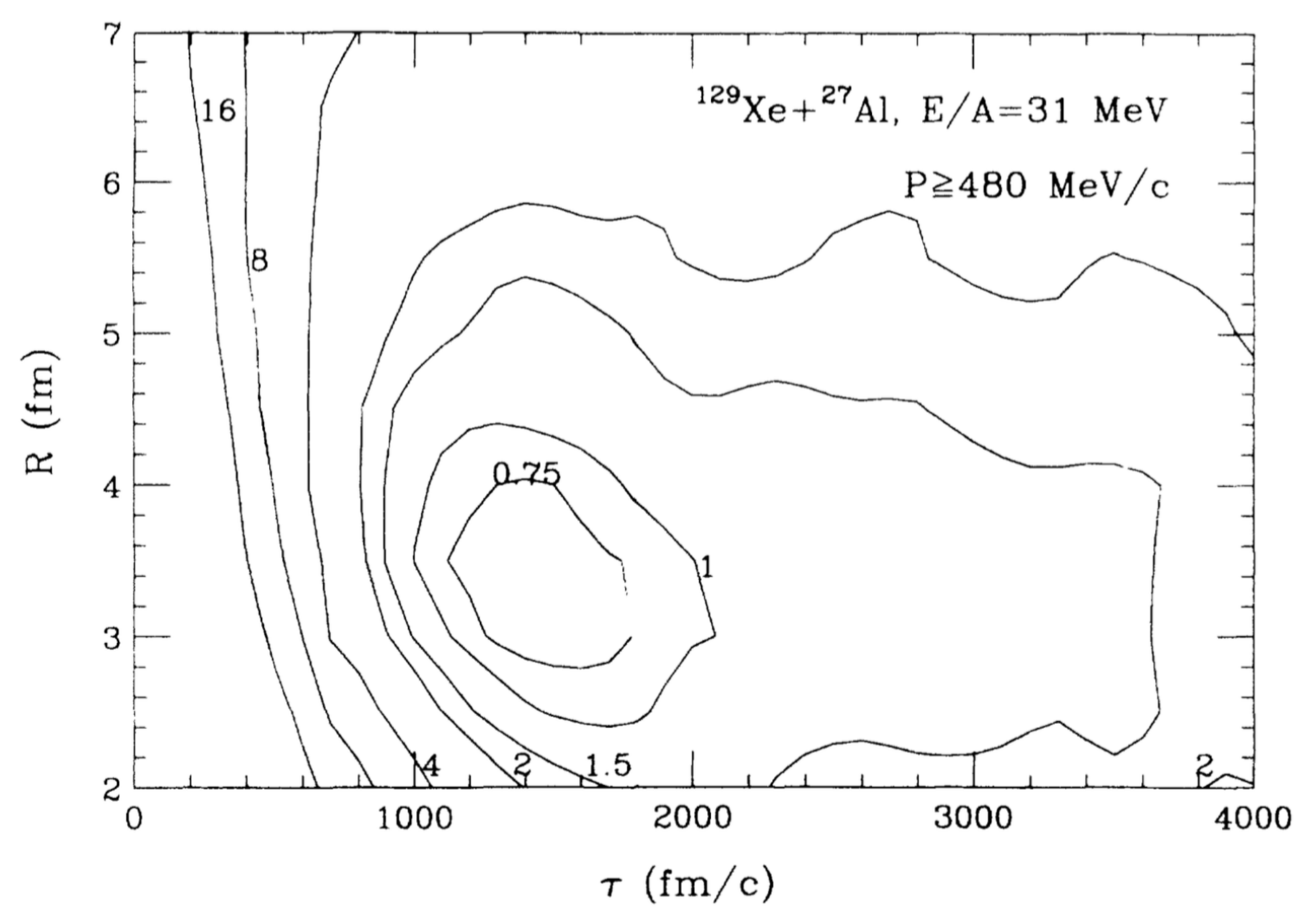}}}
\caption{
\label{fig:Lisa1994fig2}
Contour plot of $\chi^2/\nu$ evaluated by comparing measured longitudinal and transverse
  correlation functions (over the range $15~{\rm fm/c}\leq q\leq40~{\rm fm/c}$ to those
  predicted for emission from a schematic source with radius and lifetime parameters $R$ and $\tau$.
From~\cite{Lisa:1994zz}.
}
\end{figure}

Figure~\ref{fig:Lisa1994fig2} quantifies the sensitivity of the parameter extraction, through
  contours of the chi-square per degree of freedom, analogous to that of figure~\ref{fig:Lisa1993fig3}.

This result is virtually unknown in the relativistic heavy ion community, which is unfortunate-- 1300~fm/c!
This value, which is precisely what one expects for an evaporating compound nucleus at this energy,
  remains the longest timescale ever measured with multidimensional intensity interferometry in subatomic physics.

\section{Evidence for a burning log}

There have long been predictions~\cite{Pratt:1986cc,Bertsch:1989vn,Rischke:1996em} that a first-order transition
  from a deconfined state (quark-gluon plasma) to a confined (hadronic) final state, may lead to an increase
  in the system emission time.
The expectation~\cite{Rischke:1996em} is that this increase should occur just at the threshold energy
  for which a deconfined state is formed.
At lower energies there is no transition at all, whereas at higher energies the system is exploding too
  quickly to form a ``burning log'' scenario.
The threshold energy samples the ``softest point'' in the QCD equation of state.

At the relativistic collision energies where this phenomenon might occur, studies have used multi-dimensional pion
  interferometry~\cite{Lisa:2005dd}, where the relative momentum components (and corresponding ``HBT radii'') are
  identified in the ``Bertsch-Pratt'' decomposition~\cite{Pratt:1986cc,Bertsch:1989vn}.
Referring to figure~\ref{fig:Lisa1993fig1}, $R_{\rm out}$ measures the length scale of the pion
  cloud in the direction of the particle motion, and $R_{\rm side}$ quantifies the length scale
  perpendicular to this motion.
(At relativistic energies, where the dynamics in the beam direction are substantially different from those in the
  transverse direction, $R_{\rm out}$ and $R_{\rm side}$ are forced to be perpendicular to the beam direction, and
  a third radius, $R_{\rm long}$ quantifies the length scale along the beam.
At the lower energies discussed in sections~\ref{sec:ArSc80} and~\ref{sec:XeAl31}, where compound nucleus formation
  occurs, this distinction is not made in the ``longitudinal-transverse'' decomposition.)
In the hypothetical case where the system is not flowing, these radii are related to the emission duration
  $\tau_{\rm emission}$ as
\begin{equation}
\label{eq:LifetimeApprox}
R_{\rm out}^2 \approx R_{\rm side}^2 + \beta^2\tau_{\rm emission}^2 ,
\end{equation}
where $\beta=p_{\perp}/E$ is the pion speed in the transverse direction.
Relativistic heavy ion collisions, however, are dominated by transverse flow,
  so equation~\ref{eq:LifetimeApprox} is only a crude approximation~\cite{Retiere:2003kf}; indeed, $R_{\rm out}$ can be less than
  $R_{\rm side}$ at high $p_T$~\cite{Adamczyk:2014mxp}.

\begin{figure}[t]
{\centerline{\includegraphics[width=0.9\textwidth]{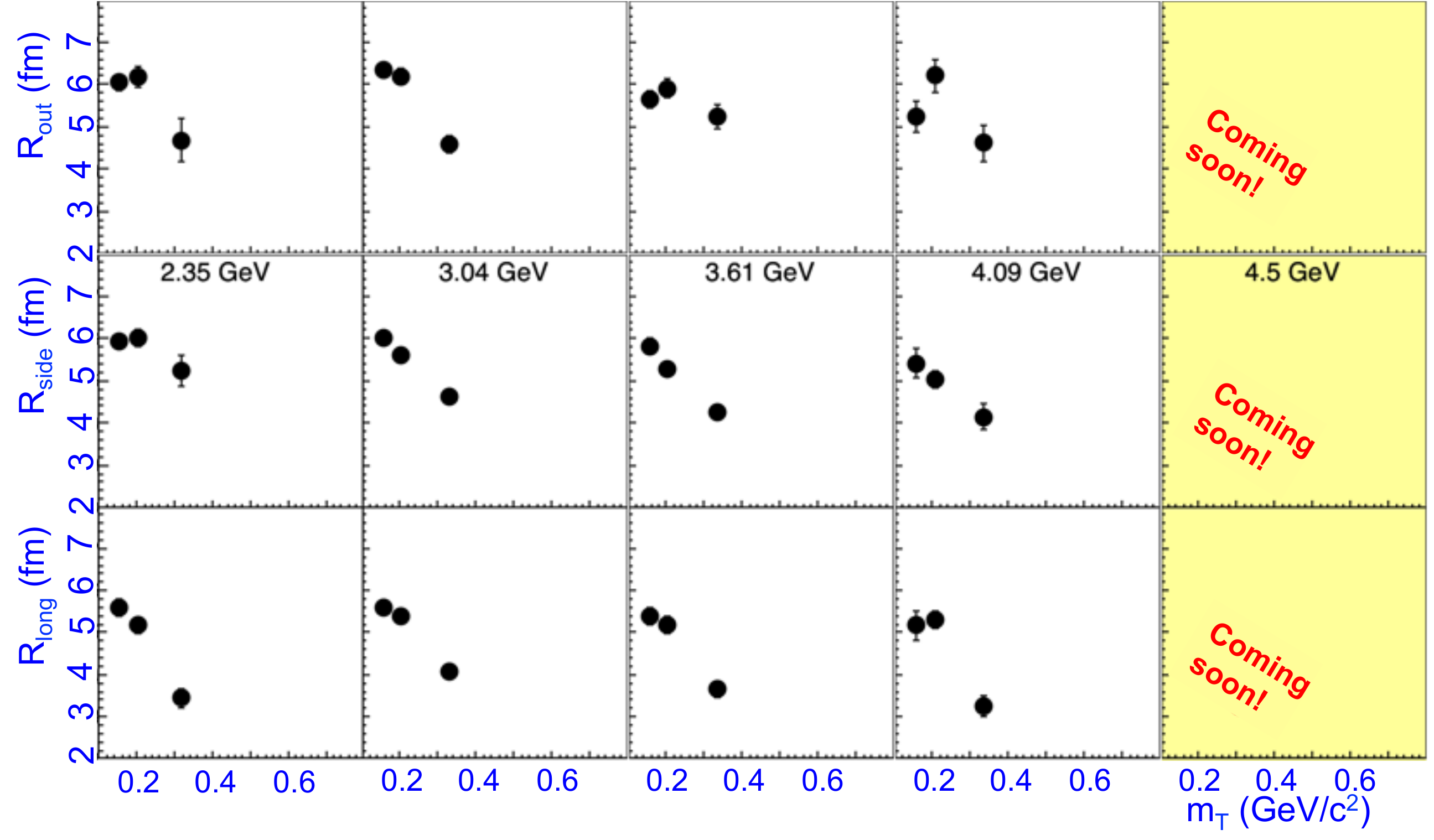}}}
{\centerline{\includegraphics[width=0.9\textwidth]{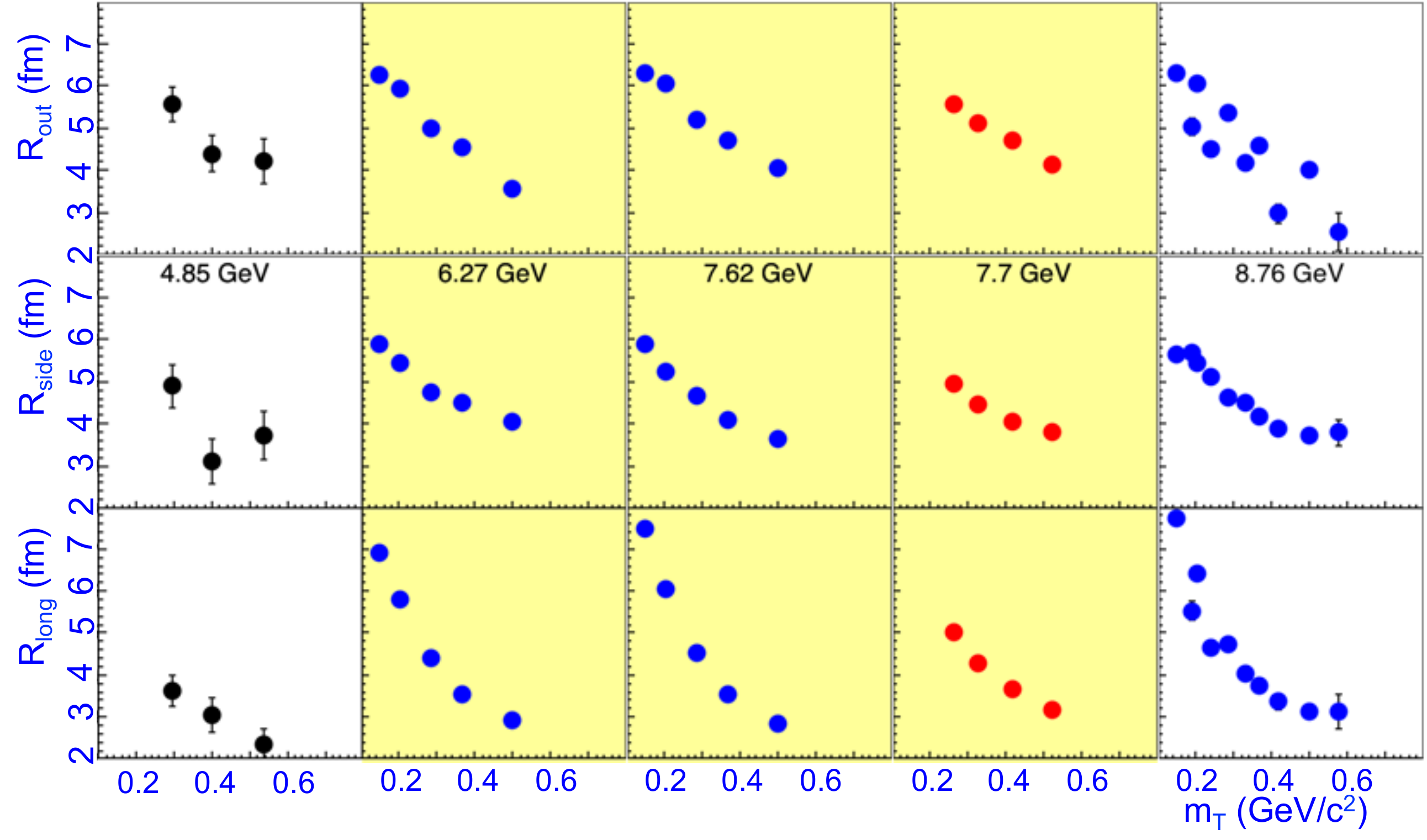}}}
\caption{
\label{fig:AllPionData1}
Two-pion femtoscopy has been measured in central heavy ion collisions over three orders
  of magnitude.
Above, HBT radii from $\sqrt{s_{NN}}=2.35-8.76$~GeV collisions are plotted versus the transverse
  mass of the pair.
Figure~\ref{fig:AllPionData2} shows analogous data up to $\sqrt{s_{NN}}=2760$~GeV.\newline
Black datapoints originate from experiments at the AGS;
  red datapoints originate from experiments at RHIC;
  blue datapoints originate from experiments at the SPS;
  pink datapoints originate from experiments at the LHC.\newline
Yellow panels identify measurements done after a 2005 review~\cite{Lisa:2005dd}.}
\end{figure}

\begin{figure}[t]
{\centerline{\includegraphics[width=0.9\textwidth]{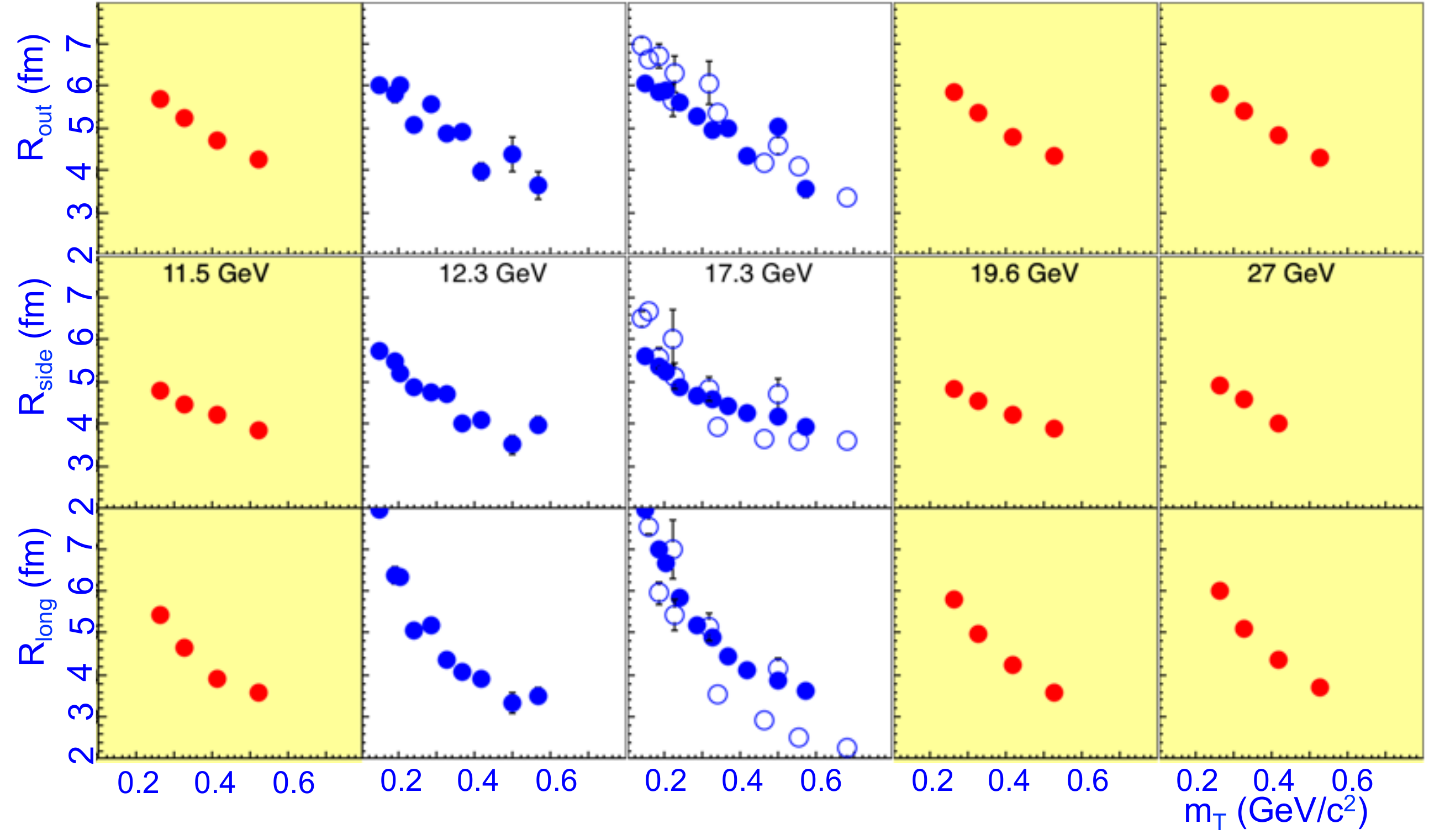}}}
{\centerline{\includegraphics[width=0.9\textwidth]{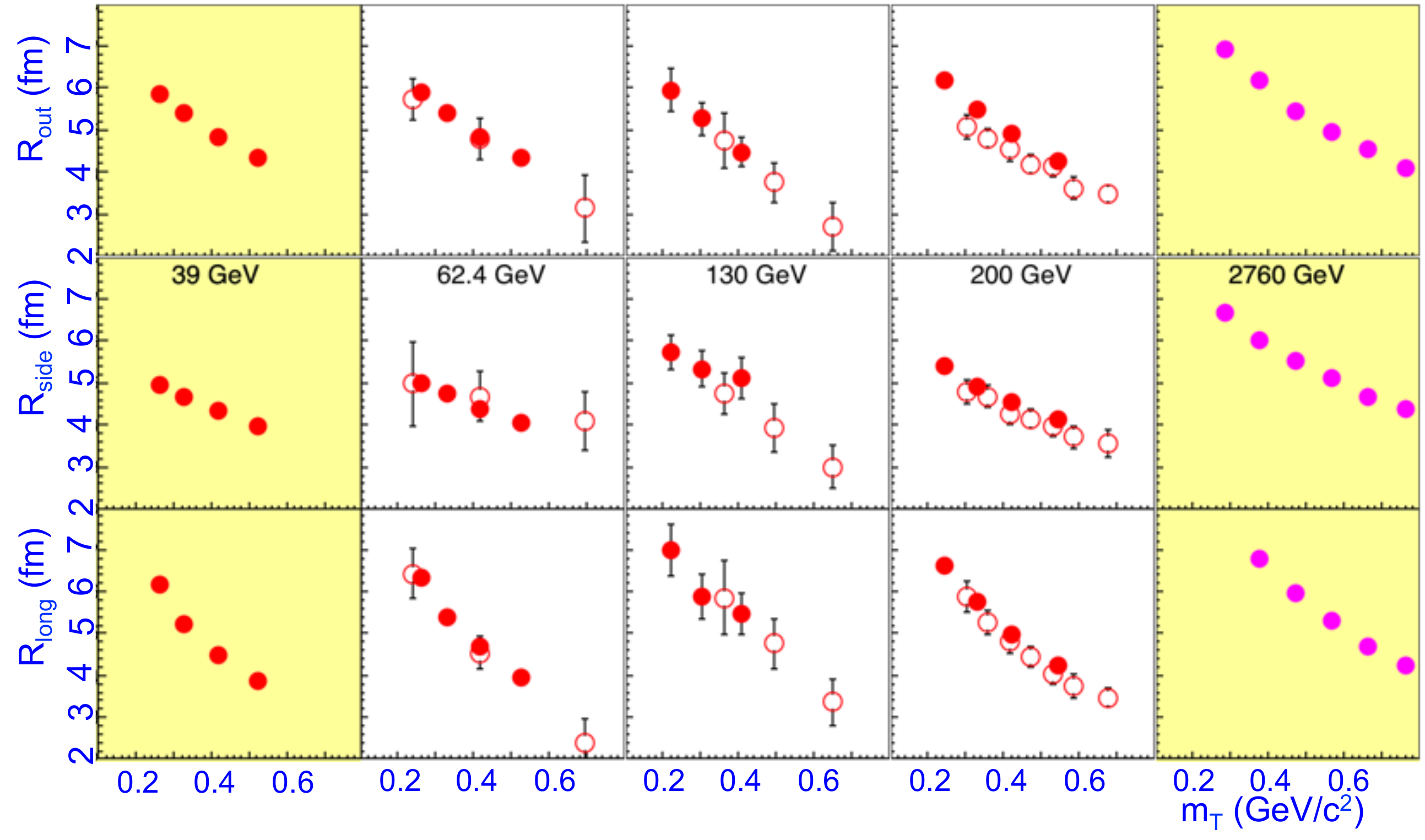}}}
\caption{
\label{fig:AllPionData2}
Two-pion femtoscopy has been measured in central heavy ion collisions over three orders
  of magnitude.
Above, HBT radii from $\sqrt{s_{NN}}=11.5-2760$~GeV collisions are plotted versus the transverse
  mass of the pair.
Figure~\ref{fig:AllPionData1} shows analogous data down to $\sqrt{s_{NN}}=2.35$~GeV.\newline
Black datapoints originate from experiments at the AGS;
  red datapoints originate from experiments at RHIC;
  blue datapoints originate from experiments at the SPS;
  pink datapoints originate from experiments at the LHC.\newline
Yellow panels identify measurements done after a 2005 review~\cite{Lisa:2005dd}.}
\end{figure}

A review of femtoscopic results in 2005~\cite{Lisa:2005dd} concluded that there was no evidence for the
  burning log signature in the two decades of relativistic heavy ion measurements at the AGS, SPS and RHIC.
Since that review, another decade has passed, and many more measurements have been done.
Figures~\ref{fig:AllPionData1} and~\ref{fig:AllPionData2} contain the world dataset of pion HBT radii
  from collisions of the heaviest nuclei (Au+Au in the U.S. and Pb+Pb in Europe).

Datapoints in the yellow panels correspond to measurements that were performed in last decade.
ALICE measurements~\cite{Adam:2015vna} at the LHC extend the measured energy range to three
  orders of magnitude.
More important, however, are the measurements at RHIC and the SPS at energies {\it below} the maximum
  energy of the machine.
These data were taken in ``energy scan'' programs, motivated by the increasing realization that
  some of the most important phenomena in hot QCD physics might only be revealed by a careful, systematic
  study of heavy ion collisions as the system conditions are gradually changed.

The versatility of the RHIC collider is clear from the fact that the RHIC data (red data points) extend
  to low energies well below traditional SPS energy of 17.3~GeV.
STAR has collected data in ``fixed-target mode,'' in which one low-energy RHIC beam struck a gold
  foil placed toward the edge of the beam pipe at one end of the STAR detector.
Despite the fact that STAR is designed for midrapidity measurements at a 200-GeV collider, the
  data taken were good, and HBT radii, fully in line with data at similar energies, have been
  measured.
At this moment, these results are unavailable for release; however, they are firm, and 
  I could not resist a placeholder in figure~\ref{fig:AllPionData1} indicating that RHIC has
  now extended measurements into the AGS energy range.

\begin{figure}[t]
{\centerline{\includegraphics[width=0.9\textwidth]{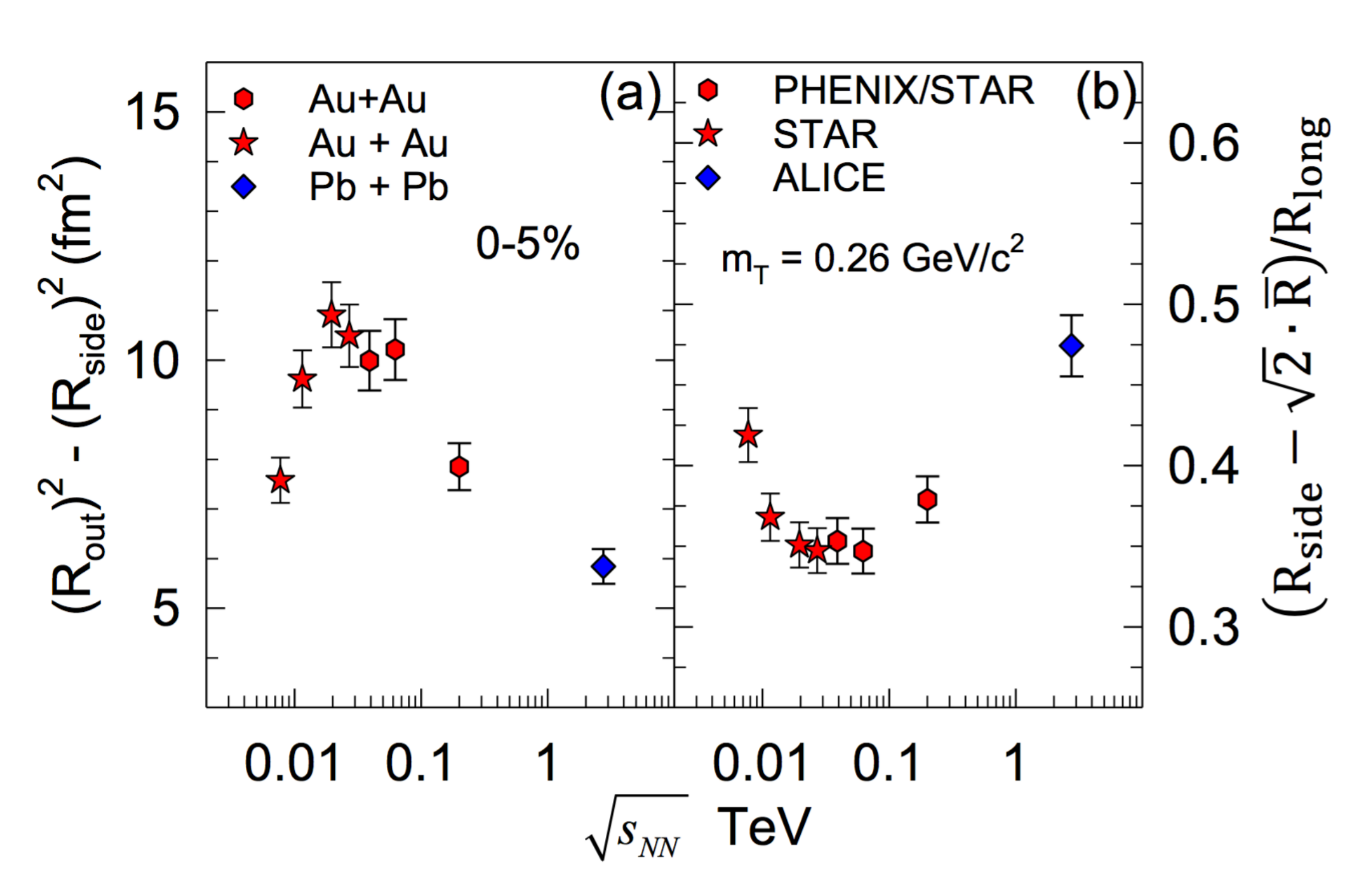}}}
\caption{
\label{fig:PHENIXBurningLog}
The difference (or ratio) of $R_{\rm out}$ and $R_{\rm side}$ is related to the emission duration
  of the collision.
As discussed in the text, a generic expectation from a first-order phase transition is a rise and
  fall of this difference, with collision energy.
Data from the RHIC Beam Energy Scan appears to validate this prediction.
Figure from~\cite{Adare:2014qvs}.
}
\end{figure}

The energy scan at RHIC may have finally revealed the burning log signature, as shown in figure~\ref{fig:PHENIXBurningLog}.
A clear peak in $R_{\rm out}^2-R_{\rm side}^2$ (or $R_{\rm out}/R_{\rm side}$~\cite{Adamczyk:2014mxp}) is observed
  around $\sqrt{s_{NN}}=15$~GeV, an energy region where other intriguing phenomena have been reported~\cite{Adamczyk:2013dal,Adamczyk:2014ipa}.
This figure includes only data from RHIC and LHC collider experiments; these have all been performed with
  similar techniques and acceptances.
Experiments at the CERN SPS have acceptances which vary with $\sqrt{s_{NN}}$, making them not ideal
  for searching for subtle changes as collision energy changes; femtoscopic results fluctuate significantly
  and disagree experiment-to-experiment.
Furthermore, SPS measurements are performed with a variety of methods to handle the Coulomb effect; this
  can affect HBT radii significantly~\cite{Adams:2004yc}.
RHIC and LHC experiments all use the so-called Bowler-Sinyukov~\cite{Bowler:1991vx,Bowler:1987it} approach, explicitly including Coulomb
  effects in the fits to the correlation functions.

It is increasingly important that hydrodynamic theory address the RHIC Beam Energy Scan range in detail.
While calculations at LHC energies are technically easier to perform (due to approximate boost invariance, a simple Equation of State,
  and low viscosity), the lower energies are more important.
QCD has a scale, after all.
Just as solid state physicists study superconductivity around the transition point, heavy ion studies must focus
  on the energy region set by QCD physics.

\section{The evolution time of the system}

Thus far, I have discussed measurements of the emission duration $\tau_{\rm emission}$ of the hot system generated in
  a heavy ion collision.
An estimate of the evolution timescale, $\tau_{\rm evolution}$, is also crucial for a detailed understanding of
  the system's dynamics.

\subsection{Estimate based on the longitudinal radius}
\label{sec:EvolutionTimeRlong}

At the very low energies discussed in sections~\ref{sec:ArSc80} and~\ref{sec:XeAl31}, it is unclear how to
  distinguish the system evolution time from the emission duration.
However, as Sinyukov and collaborators pointed out~\cite{Makhlin:1987gm},
  in ultrarelativistic collisions, the strong longitudinal flow
  generates a nearly boost-invariant system in which the longitudinal HBT radius and evolution time
  are related by~\cite{Makhlin:1987gm,Wiedemann:1995au}
\begin{equation}
\label{eq:Sinyukov}
R_{\rm long}^2\left(m_T\right) \approx \tau_{\rm evolution}^2\frac{T}{m_T}\frac{{\rm K}_2\left(m_T/T\right)}{{\rm K}_1\left(m_T/T\right)} ,
\end{equation}
where $T$ is the system temperature at freezeout, and $m_T$ is the transverse mass of the particles.

\begin{figure}[t]
{\centerline{\includegraphics[width=0.7\textwidth]{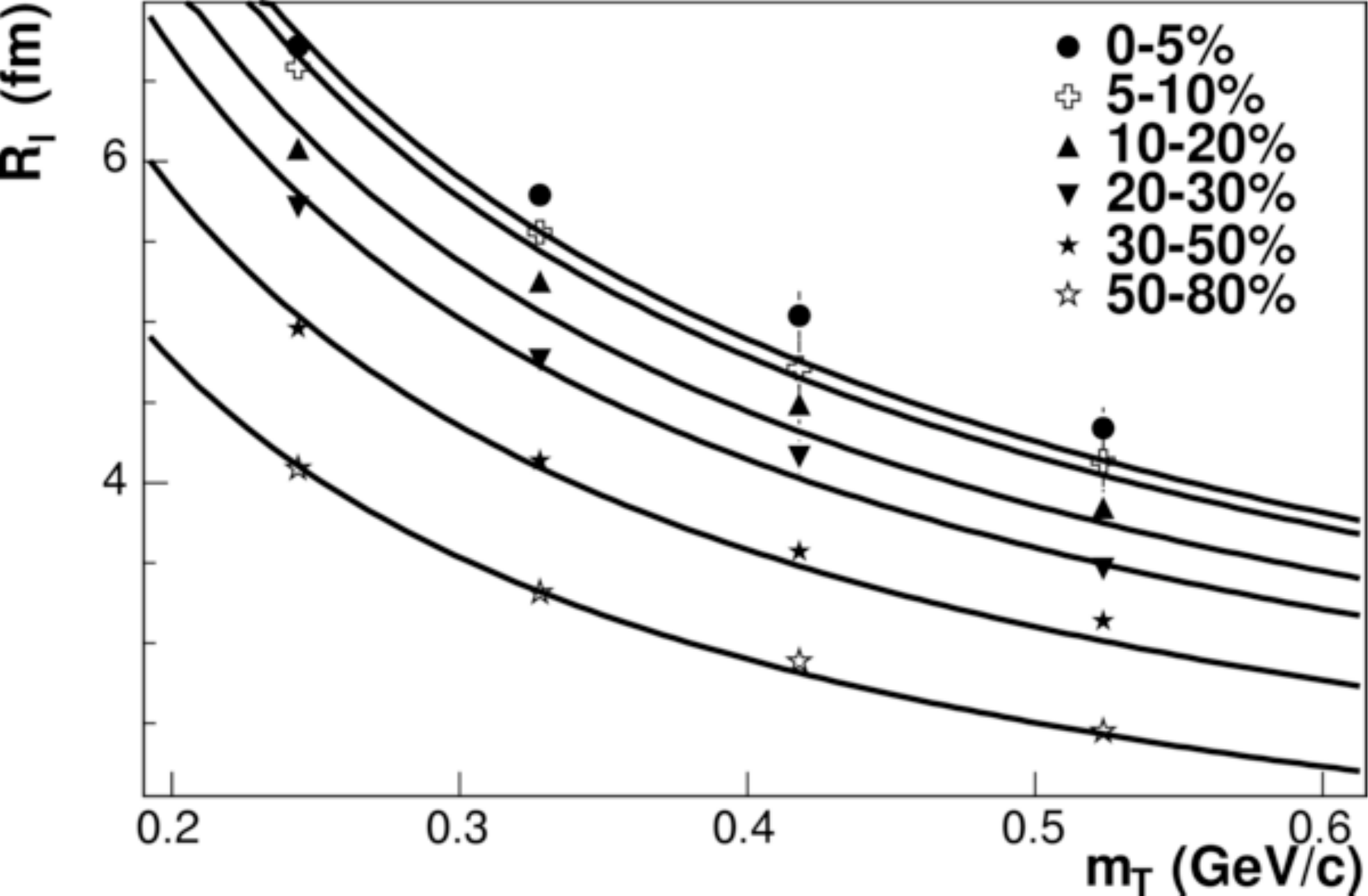}}}
\caption{
\label{fig:RlongMercedes}
Longitudinal HBT radii for $\sqrt{s_{NN}}=200$~GeV Au+Au collisions of varying centrality, measured by the STAR Collaboration.
From~\cite{Adams:2004yc}.
}
\end{figure}

\begin{figure}[t]
{\centerline{\includegraphics[width=0.5\textwidth]{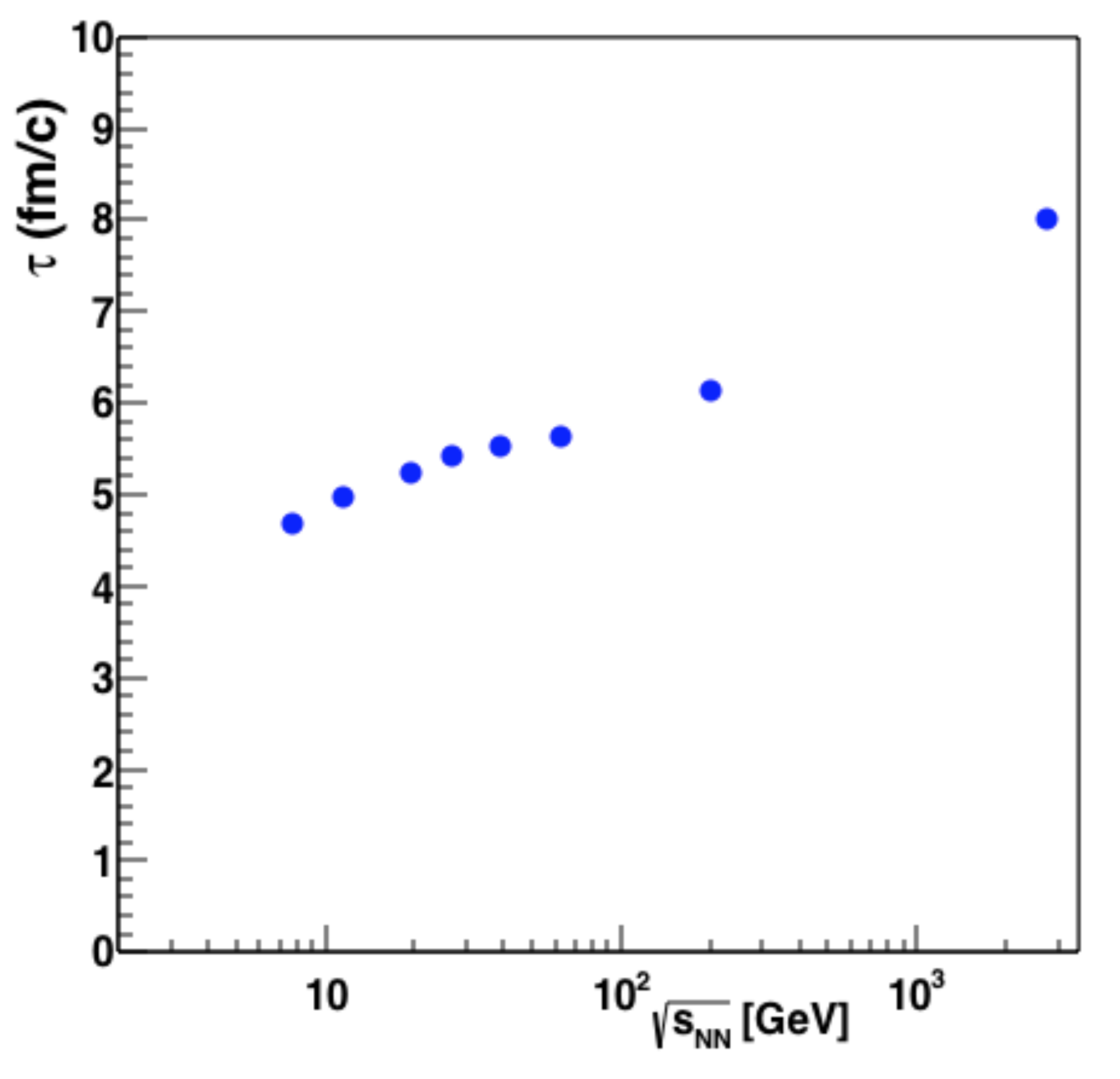}}}
\caption{
\label{fig:EvolutionTimes}
Estimate of the system evolution time based on Sinyukov fits (equation~\ref{eq:Sinyukov}) to measured longitudinal HBT radii $R_{\rm long}$.
}
\end{figure}

Figure~\ref{fig:RlongMercedes} shows fits of formula~\ref{eq:Sinyukov} to longitudinal radii measured~\cite{Adams:2004yc}
  by the STAR Collaboration for collisions at $\sqrt{s_{NN}}=200$~GeV at various centralities.
The fit is reasonable.
Evolution timescales extracted from STAR~\cite{Adamczyk:2014mxp} and ALICE~\cite{Adam:2015vna} are shown in figure~\ref{fig:EvolutionTimes}.

\subsection{Alternate cross-check of the evolution time estimate}

In section~\ref{sec:EvolutionTimeRlong}, I outlined the ``traditional'' way to estimate the evolution time,
  based on the $m_T$ dependence of the longitudinal HBT radius.
Here, I provide an independent cross-check from another direction.

In non-central heavy ion collisions, the hot system is initially anisotropic relative to the reaction plane (spanned by
  the impact parameter vector and the beam direction) of the collision.
The response of the system to this coordinate-space anisotropy generates a corresponding momentum-space anisotropy,
  in which more (and faster) particles are emitted in the reaction plane than perpendicular to it.
This is the well-known ``elliptic flow'' phenomenon, often quantified by a momentum-space anisotropy parameter $v_2$~\cite{Voloshin:2008dg}.

\begin{figure}[t]
{\centerline{\includegraphics[width=0.6\textwidth]{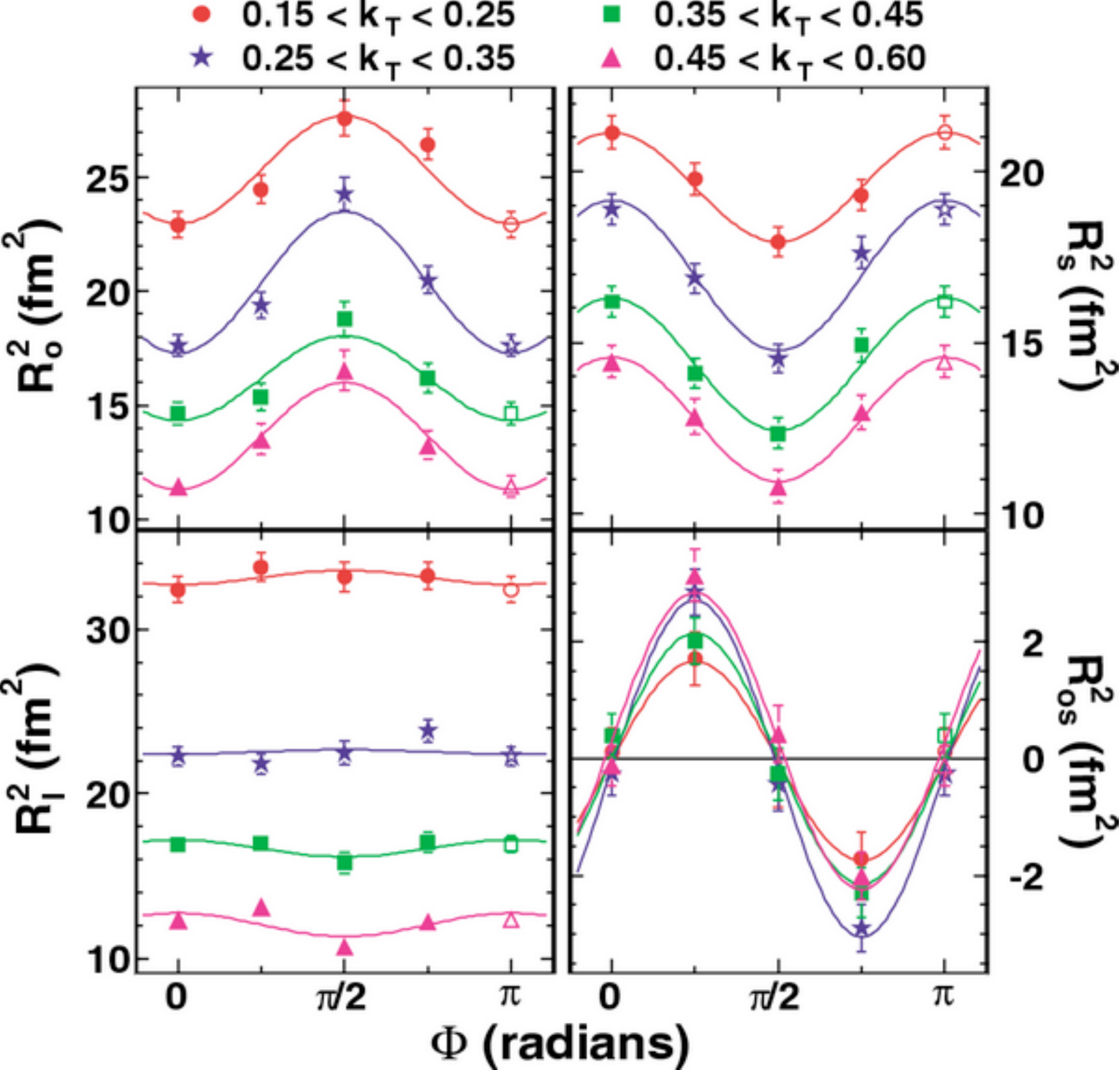}}}
\caption{
\label{fig:STARasHBT}
Pion HBT radii from mid-central Au+Au collisions at $\sqrt{s_{NN}}=200$~GeV, as a function
  of azimuthal angle relative to the event plane.
From~\cite{Adams:2003ra}.
}
\end{figure}

The preferentially in-plane expansion will tend to reduce (or perhaps reverse) the anisotropy of the initial state;
  i.e. the system will become more round in coordinate space.
If the system retains some anisotropy, the transverse HBT radii will oscillate as a function of azimuthal angle relative
  to the reaction plane~\cite{Lisa:2000ip,Heinz:2002au,Retiere:2003kf}.
Figure~\ref{fig:STARasHBT} shows femtoscopic radii measured~\cite{Adams:2003ra} by the STAR Collaboration for mid-central
  Au+Au collisions at $\sqrt{s_{NN}}=200$~GeV.
Fourier coefficients, the space-time analog to $v_2$, may be extracted from the oscillations and used to estimate the
  spatial anisotropy of the source at freeze-out~\cite{Retiere:2003kf}, defined as
\begin{equation}
\label{eq:epsilon}
\epsilon_{\rm F}\equiv\frac{\sigma_y^2-\sigma_x^2}{\sigma_y^2+\sigma_x^2} \quad,
\end{equation}
where $\sigma_x$ ($\sigma_y$) is the root-mean-square extent of the source in (out of)
  the event plane.

\begin{figure}[t]
{\centerline{\includegraphics[width=1.0\textwidth]{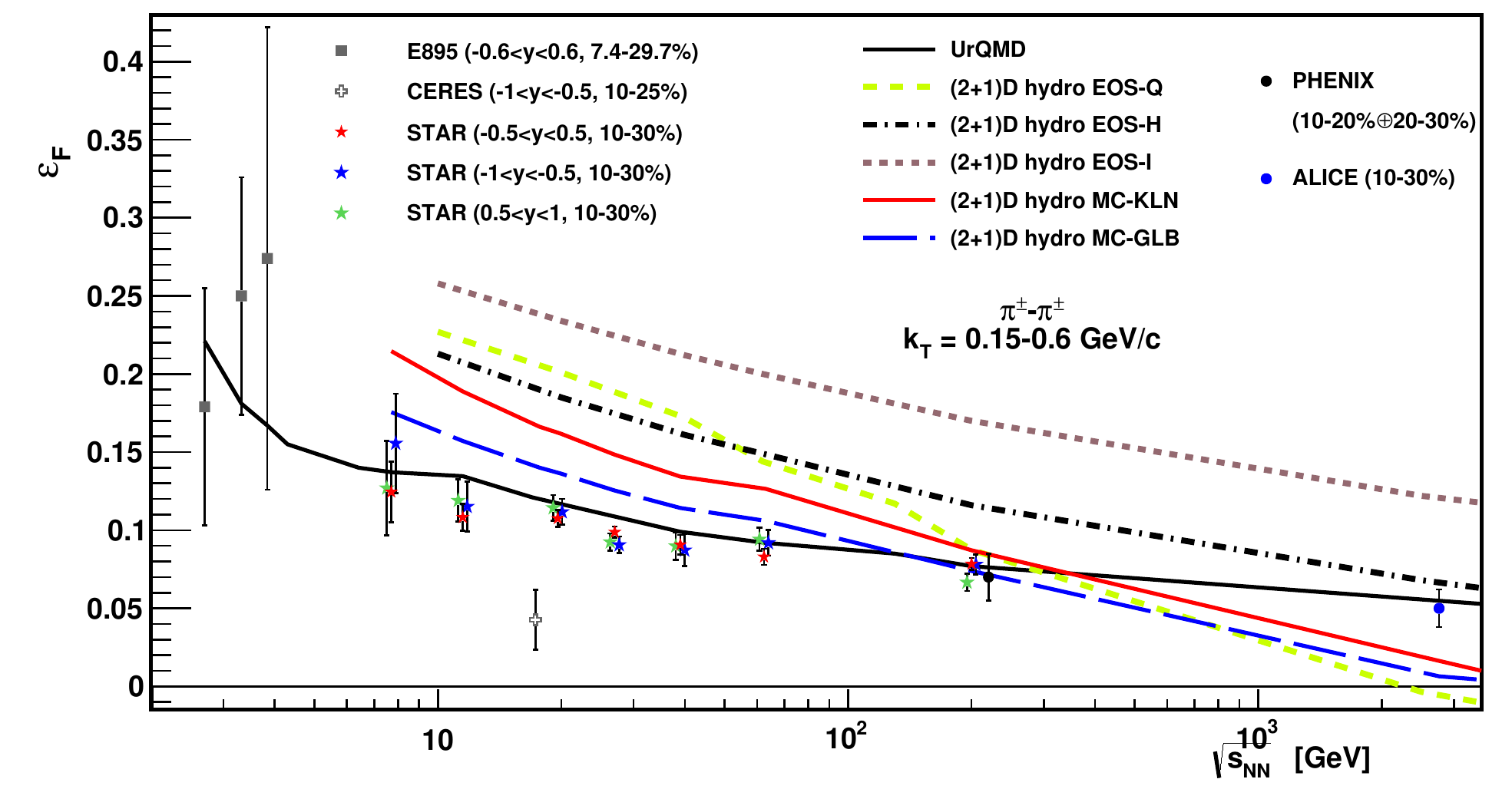}}}
\caption{
\label{fig:AllAsHbt}
The spatial anisotropy, defined in equation~\ref{eq:epsilon}, for mid-central Au+Au (Pb+Pb) collisions
  from E895/AGS~\cite{Lisa:2000ip}, CERES/SPS~\cite{Adamova:2008hs}, STAR/RHIC~\cite{Adams:2003ra,Adamczyk:2014mxp}, 
  PHENIX/RHIC~\cite{Adare:2014vax} and ALICE/LHC~\cite{Loggins:2014uaa}.
Calculations~\cite{Kolb:2003dz,Lisa:2011na,Shen:2012vn} with hydrodynamic and transport models are shown for comparison.
}
\end{figure}

Spatial anisotropies have been extracted for heavy ion collisions over the entire available energy range and
  are plotted in figure~\ref{fig:AllAsHbt}.
For all energies, the system retains its out-of-plane extension in coordinate space, though it is reduced
  from its initial value of 0.25 (estimated from Glauber calculations).

The evolution of the hot system produced in a heavy ion collision is complex, but toy models can be useful
  to check whether disparate measurements may be understood in a single simple scenario.
In this spirit, I construct a blast-wave~\cite{Retiere:2003kf} inspired model and
  ask whether the evolution times plotted in figure~\ref{fig:EvolutionTimes} are reasonably
  consistent with the reduction in coordinate-space anisotropy seen in figure~\ref{fig:AllAsHbt}.

Let $\overline{\sigma}_0$ be the angle-averaged RMS size of the source at $t=0$, and $\epsilon_0$ be its anisotropy.
Further, let $\beta_x$ and $\beta_y$ be the average flow velocities in and out of the reaction plane.
Assuming constant anisotropic free-streaming evolution of the source, its spatial anisotropy after 
  evolving for $\tau_{\rm evolution}$ is
\begin{equation}
\label{eq:ToyModel}
\epsilon_{\rm F}\left(\tau_{\rm evolution}\right) = 
   \frac{\overline{\sigma}_0^2\epsilon_0-\tfrac{1}{2}\left(\beta_y^2-\beta_x^2\right)\tau_{\rm evolution}^2}{\overline{\sigma}_0^2\epsilon_0+\tfrac{1}{2}\left(\beta_y^2+\beta_x^2\right)\tau_{\rm evolution}^2}
\end{equation}

Based on Glauber calculations, $\sigma_0\approx3.5$~fm and $\epsilon_0\approx0.25$.
The evolution time $\tau_{\rm evolution}$ for the 10-30\% central collisions were extracted from $R_{\rm long}$ for that centrality.
Average transverse flow velocities are related to blast-wave~\cite{Retiere:2003kf} flow parameters according to
\begin{equation*}
\beta_y = \tanh\left(2\left(\rho_0-\rho_2\right)/3\right) \qquad
\beta_x = \tanh\left(2\left(\rho_0+\rho_2\right)/3\right) .
\end{equation*}
Using blast-wave fit parameters extracted by the PHENIX collaboration~\cite{Adare:2015bcj},
  $\beta_x=0.585$ and $\beta_y=0.490$.
While one expects higher flow velocities at higher energies, the same values for $\beta_x$ and $\beta_y$ were used for all $\sqrt{s_{NN}}$,
  since other blast-wave fits were not readily unavailable.
However, it turns out that these velocities vary little with $\sqrt{s_{NN}}$, so these values should serve for a test.

\begin{figure}[t]
{\centerline{\includegraphics[width=1.0\textwidth]{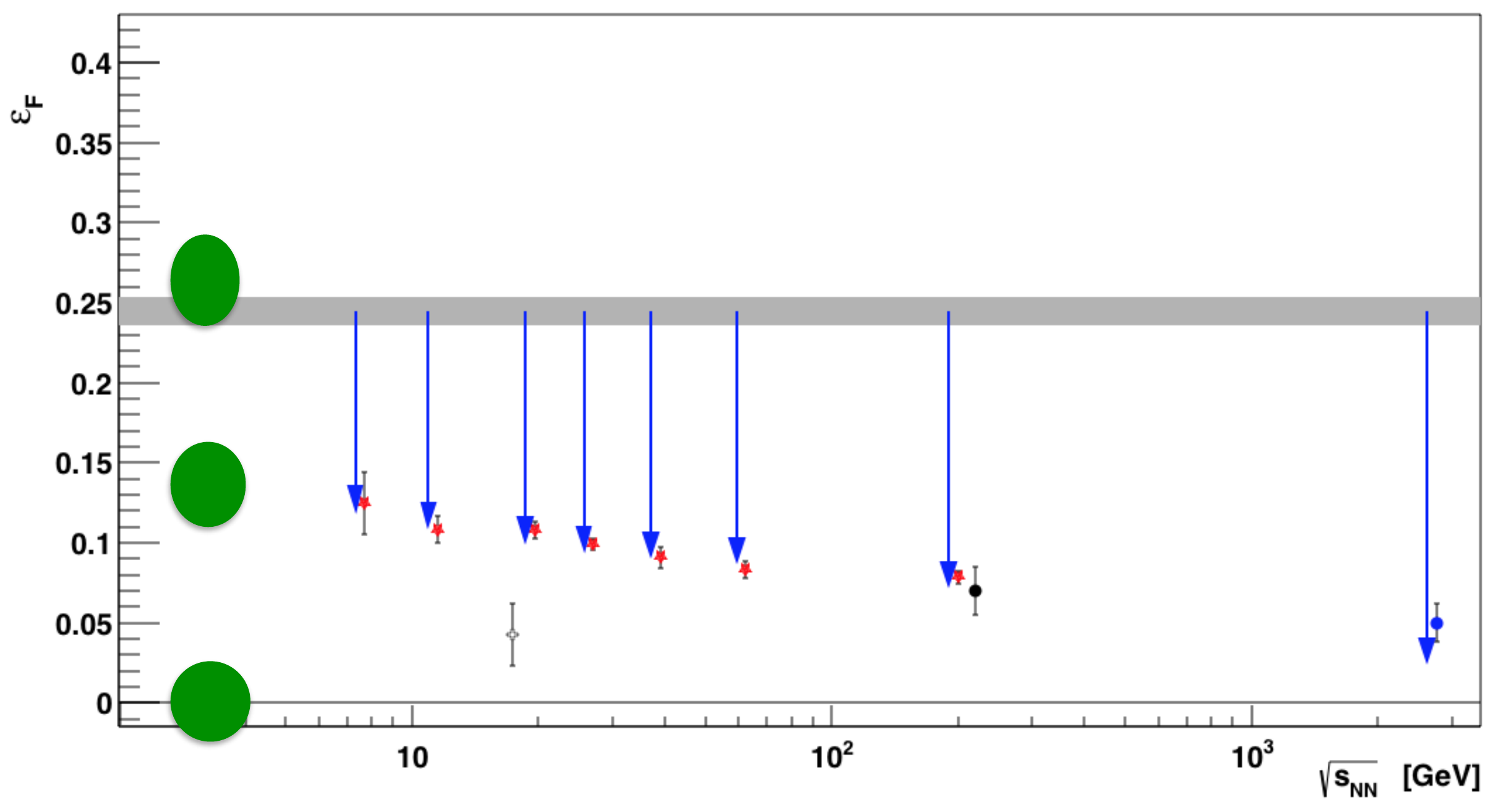}}}
\caption{
\label{fig:ToyModel}
Measurements of the final freeze-out eccentricity (from figure~\ref{fig:AllAsHbt} are compared with 
  calculations of a toy model based on an initially out-of-plane extended source evolving with preferential in-plane expansion.
The grey band indicates the initial anisotropy based on Glauber calculations, and the downward-facing arrows indicate the
  evolution of the shape.
The terminus of the arrow corresponds to the shape size at the time $\tau_{\rm evolution}$ extracted by Sinyukov fits (equation~\ref{eq:Sinyukov}
  to the longitudinal radii.
See equation~\ref{eq:ToyModel} and text for details.
}
\end{figure}

The estimate from this toy model is compared to data in figure~\ref{fig:ToyModel}.
Considering its crudity and not tinkering with parameters, the agreement is remarkable.
Both the magnitude and the $\sqrt{s_{NN}}$-dependence of $\epsilon_{\rm F}$ seem to be consistent with an evolution timescale
  extracted in the ``traditional'' way, using $R_{\rm long}$, as discussed in section~\ref{sec:EvolutionTimeRlong}.

\section{Summary}

To understand the dynamics of a heavy ion collision, it is important to have an estimate
  of the timescales associated with its evolution.
I have discussed experimental measurements of the emission and evolution timescales based
  on particle intensity interferometry measurements.

When the analysis was performed in the correct reference frame, two-proton correlation functions
  at low collision energies revealed long lifetimes, consistent with theoretical expectations.
These observations were important, as they put to rest troubling doubts about our understanding
  of intensity interferometry overall.

In ultra-relativistic heavy ion collisions, the long-sought ``burning log'' signature of a softening
  of the equation of state was found, but only after a systematic scan of the collision energy.
This is one of several interesting signatures at energies around $\sqrt{s_{NN}}\approx20$~GeV that
  have been revealed in the RHIC Beam Energy Scan program.

The evolution time of a collision is found to grow with collision energy, and the traditional estimate
  based on the longitudinal HBT radius was found to be consistent with a toy model describing the evolution
  of the spatially anisotropic source as estimated by azimuthal oscillations of the transverse HBT radii.

These timescale estimates serve as important input to theoretical studies of the dynamics of the collision.
Such studies are crucial, if the field is to generate lasting physics contributions to our understanding of QCD.
While dynamic modeling of the highest-energy collisions (e.g. at the LHC) are much easier, it is much 
  more important to focus on lower energies around $\sqrt{s_{NN}}\sim20$~GeV, where nontrivial phenomena
  associated with the QCD equation of state may appear.

Finally, I would like to congratulate Prof. Andrzej Bia\l as on the occasion of his $80^{\rm th}$ birthday,
  from one ant to another.



\end{document}